\documentclass[letterpaper, 10 pt, conference]{ieeeconf}
\IEEEoverridecommandlockouts
\let\labelindent\relax
\usepackage{enumitem}
\usepackage{balance}
\usepackage[font={small}]{caption}
\usepackage{subcaption}
\usepackage{array}
\usepackage{textcomp}
\usepackage{mathtools, nccmath}
\usepackage{graphicx}
\usepackage{amsfonts}
\usepackage{amsmath}
\usepackage{amssymb}
\usepackage{algorithm}
\usepackage{algorithmic}
\usepackage{hyperref}
\usepackage{tikz}
\usepackage{arydshln}
\usepackage{multirow}
\usepackage{bm}
\usepackage{epstopdf}
\usepackage{cite}
\usepackage{siunitx}

\DeclareMathOperator*{\argmin}{argmin} 

\usepackage{amsthm}

\newtheoremstyle{mystyle}
  {}
  {}
  {}
  {}
  {\bfseries}
  {.}
  { }
  {\thmname{#1}\thmnumber{ #2}\thmnote{ (#3)}}

\theoremstyle{mystyle}

\newtheorem{remark}{Remark}

\begin{document}
\title{\LARGE \bf Enhancing Feasibility and Safety of Nonlinear Model Predictive Control \\ with Discrete-Time Control Barrier Functions}
\author{Jun Zeng$^*$, Zhongyu Li$^*$, and Koushil Sreenath
\thanks{$^*$ Authors have contributed equally.}
\thanks{This work was partially supported through National Science Foundation Grant CMMI-1931853.}
\thanks{All authors are with Hybrid Robotics Group at the Department of Mechanical Engineering, UC Berkeley, USA. \tt\small\{zengjunsjtu, zhongyu\_li, koushils\}@berkeley.edu}
\thanks{The reference code for implementing this paper can be found at \url{https://github.com/HybridRobotics/NMPC-DCLF-DCBF}}
}
\maketitle

\begin{abstract}
Safety is one of the fundamental problems in robotics.
Recently, one-step or multi-step optimal control problems for discrete-time nonlinear dynamical system were formulated to offer tracking stability using control Lyapunov functions (CLFs) while subject to input constraints as well as safety-critical constraints using control barrier functions (CBFs).
The limitations of these existing approaches are mainly about feasibility and safety.
In the existing approaches, the feasibility of the optimization and the system safety cannot be enhanced at the same time theoretically.
In this paper, we propose two formulations that unifies CLFs and CBFs under the framework of nonlinear model predictive control (NMPC). 
In the proposed formulations, safety criteria is commonly formulated as CBF constraints and stability performance is ensured with either a terminal cost function or CLF constraints.
Slack variables with relaxing technique are introduced on the CBF constraints to resolve the tradeoff between feasibility and safety so that they can be enhanced at the same.
The advantages about feasibility and safety of proposed formulations compared with existing methods are analyzed theoretically and validated with numerical results.
\end{abstract}
\section{Introduction}
\label{sec:introduction}
\subsection{Motivation}
Safety-critical optimal control and planning is one of the fundamental problems in robotic applications.
In order to ensure the safety of robotic systems while achieving optimal performance, tight coupling between potentially conflicting control objectives and safety criteria is considered in an optimization problem.
Some researchers formulate this problem using control barrier functions under the continuous dynamics of the system \cite{ames2014control, ames2016control}, where the optimal performance is achieved by the control Lyapunov functions and safety criteria is guaranteed through control barrier functions.
Recently, this methodology is also introduced in the discrete-time domain and the optimal control problem can be formulated to calculate the current optimal input \cite{agrawal2017discrete} or a sequence of ones in the fashion of model predictive control \cite{zeng2020safety}.
However, feasibility and safety are still the main challenges for these formulations using discrete-time control barrier functions.
In this paper, our proposed formulations focus on how to handle these problems and provides a detailed analysis comparing to the existing methods using discrete-time control barrier functions.

\subsection{Related Work}
Designing controllers to ensure provable safety guarantees for autonomous systems is vital.
One approach to provide safety guarantees in control is to draw inspirations from control barrier functions \cite{ames2019control}.
The CBF-QP formulation \cite{ames2014rapidly} permits us to find the minimum perturbation for a given feedback controller to guarantee safety.
Control Lyapunonv functions (CLFs) \cite{romdlony2016stabilization} can be applied to stabilize the closed-loop dynamics of both linear and nonlinear dynamical systems \cite{galloway2015torque}.
Together with CBFs, the CLF-CBF-QP formulation \cite{ames2016control} enables handling safety-critical constraints effectively in real-time.
This approach is also generalized for high-order systems \cite{nguyen2016exponential, xiao2019control}. Robust or adaptive optimal control are also applied with this technique \cite{nguyen2016optimal, jankovic2018robust, taylor2020adaptive, lopez2020robust}.

\subsubsection{Optimal Control with Discrete-time CBFs}
Besides the continuous-time domain, the formulations of CBFs are generalized into discrete-time systems. An optimization problem can be formulated to calculate the current optimal control input, proposed in the DCLF-DCBF formulation in \cite{agrawal2017discrete}.
A type of model predictive control is also recently introduced to enhance performance.
The model predictive control with control Lyapunov functions (NMPC-DCLF) is proposed to ensure stability in \cite{grandia2020nonlinear}, where discrete-time CLF constraints are considered under nonlinear model predictive control (NMPC).
A control design (MPC-DCBF) for safety-critical tasks is firstly presented in \cite{zeng2020safety}, where the safety-critical constraints are enforced by discrete-time control barrier functions.
This approach could also be applied to a multi-layer control framework \cite{rosolia2020multi, grandia2020multi}, where the safety-critical control with discrete-time CBF serves as a mid-level controller or planner.

\subsubsection{Feasibility \& Safety}
Among the formulations in the discrete-time domain, optimal control with discrete-time CBFs also encounter feasibility issues.
The feasibility issues arise due to the potential empty intersection between the reachable set and the safe region confined by CBF constraints at each time step.
Moreover, with current approaches, there exists a tradeoff between safety performance and feasibility and they cannot be enhanced at the same time, as discussed in \cite{zeng2020safety}.
In other words, reducing the decay rate of CBF constraints increases the system safety, but comes at the possibility of infeasibility.
A potential way to partly handle the feasibility issue is to adopt the CBF constraint only on  the first time step as a one-step constraint, as presented in the formulation MPC-GCBF in \cite{ma2021feasibility}.
This approach is shown to enhance the feasibility, however, we are still under the restriction between choosing feasibility and safety, as the intersection between the reachable set and the safe region at the first step could still potentially be empty.
A soft constrained predictive safety filter with a terminal control barrier function is proposed in \cite{wabersich2021predictive}, which enhances the safety with respect to the model uncertainty.
However, the tradeoff between feasibility and safety is unsolved.

Similarly, feasibility is challenging to be guaranteed in the continuous domain.
Many existing approaches relax the CLF constraint to resolve the conflict between CLF and CBF constraints, as summarized in \cite{ames2019control}.
However, the QP-based problem could become infeasible as the CBF constraint might violate the input constraint even when we relax the CLF constraint.
In \cite{ames2016control}, a valid CBF is specifically designed for the adaptive cruise control scenario based on the system dynamics under input constraints, which could ensure the feasibility in the optimization.
This approach solves the problem specifically for this system but not for general nonlinear systems.
Recently in \cite{zeng2021feasibility}, the decay rates of CBF constraints are relaxed with optimization variables, which generally resolves the conflict between the CBF constraint and the input constraint and guarantees point-wise feasibility.

In this paper, the proposed formulations draw inspiration from the decay-rate relaxing technique for the CBF constraint in the continuous-time domain, and will be shown to be able to enhance the feasibility and the safety at the same time.
The proposed formulations are generalized for both one-step or multi-step constraints, and don't specifically require only an one-step constraint to increase the feasibility as done with MPC-GCBF \cite{ma2021feasibility}.

\subsection{Contributions}
The contributions of this paper are as follows:
\begin{itemize}
    \item We propose two control frameworks for guaranteeing stability and safety using nonlinear MPC (NMPC), where the safety criteria are considered as discrete-time CBF constraints, and stability criteria appear as CLFs formulated either as a terminal cost or discrete-time constraints.
    \item The decay rates of the discrete-time control barrier function constraints are relaxed in the optimization problem, which allow the proposed formulations to enhance the feasibility and the safety at the same time for both one-step and multi-step input optimization.
    \item The proposed formulations are shown theoretically to enhance the feasibility and safety performance compared with existing approaches, and also validated with numerical examples.
\end{itemize}

\subsection{Paper Structure}
The paper is organized as follows.
A brief background about discrete-time CLFs and CBFs is presented and the existing optimal control formulations are revisited in Sec. \ref{sec:background}.
The proposed formulations are presented in Sec. \ref{sec:formulation}, which unifies discrete-time CLF and CBF using NMPC.
The advantages about feasibility and safety compared with the state-of-the-art are illustrated theoretically in Sec. \ref{sec:advantages}.
Numerical simulations are shown in Sec. \ref{sec:examples} to validate our approach.
Finally, concluding remarks are provided in Sec. \ref{sec:conclusion}.
\section{Background}
\label{sec:background}

Having introduced the problem, we next present background on CLFs and CBFs and revisit some existing optimal control formulations using discrete-time CLFs and CBFs.

\subsection{CLFs and CBFs}
In this paper, we consider a discrete-time control system as follows,
\begin{equation}
    \mathbf{x}_{t+1} = f(\mathbf{x}_t, \mathbf{u}_t), \label{eq:discrete-time-dynamics}
\end{equation}
with $\mathbf{x} \in \mathcal{X}$ representing the system state with the control input $\mathbf{u}$ confined by admissible input set $\mathcal{U}$. For safety-critical control, we consider a set $\mathcal{C}$ defined as the superlevel set of a continuously differentiable function $h: \mathcal{X} \subset \mathbb{R}^n \rightarrow \mathbb{R}$,
\begin{equation}
    \begin{split}
        \mathcal{C} &= \{\mathbf{x} \in \mathbb{R}^n : h(\mathbf{x}) \geq 0 \}, \\
        \partial \mathcal{C} &= \{\mathbf{x} \in \mathbb{R}^n : h(\mathbf{x}) = 0 \}, \\
        \text{Int}(\mathcal{C}) &= \{\mathbf{x} \in \mathbb{R}^n : h(\mathbf{x}) > 0 \}.
    \end{split}
    \label{eq:cbf-safeset}
\end{equation}
Throughout this paper, we refer to $\mathcal{C}$ as a safe set.
The safe set can be regarded as the ensemble of states satisfying distance constraints
\begin{equation}
    h(\mathbf{x}) \geq 0.
    \label{eq:distance-constraint}
\end{equation}
In a stricter manner, the function $h$ becomes a control barrier function in the discrete-time domain if it satisfies the following relation,
\begin{equation}
    \Delta h(\mathbf{x}_k, \mathbf{u}_k) \geq -\gamma_k h(\mathbf{x}_k), ~ 0 < \gamma_k \leq 1, \label{eq:cbf-definition}
\end{equation}
where $\Delta h(\mathbf{x}_k, \mathbf{u}_k) := h(\mathbf{x}_{k+1})-h(\mathbf{x}_k)$. Satisfying constraint \eqref{eq:cbf-definition}, we have $h(\mathbf{x}_{k+1}) \geq (1-\gamma_k) h(\mathbf{x}_k)$, i.e, the lower bound of control barrier function $h(\mathbf{x})$ decreases exponentially at time $k$ with the rate $1-\gamma_k$.

Besides the system safety, we are also interested in stabilizing the system with a feedback control law $\mathbf{u}$ under a control Lyapunov function $V$ in the discrete-time domain,
\begin{equation}
    \Delta V(\mathbf{x}_k, \mathbf{u}_k) \leq -\alpha_k V(\mathbf{x}_k), ~0 < \alpha_k \leq 1,
\end{equation}
where $\Delta V(\mathbf{x}_k, \mathbf{u}_k) := V(\mathbf{x}_{k+1}) - V(\mathbf{x}_k)$. Similarly as above, the upper bound of control Lyapunov function decreases exponentially at time $k$ with the rate $1-\alpha_k$.

\subsection{Existing Approaches Revisited}
In this section, we will revisit some existing optimal control formulations using discrete-time CLFs or CBFs.

\subsubsection{DCLF-DCBF~\cite{agrawal2017discrete}}
The discrete-time control Lyapunov function and control barrier function can be unified into one optimization program, which achieves the control objective and guarantees system safety. This formulation was introduced in \cite{agrawal2017discrete} and is presented as follows,

\noindent\rule{\columnwidth}{0.4pt}
\textbf{DCLF-DCBF:}
\begin{subequations}
\label{eq:dclf-dcbf}
\begin{align}
    \mathbf{u}_k^* = \argmin_{(\mathbf{u}_k, s) \in \mathbb{R}^{m+1}} \quad & \mathbf{u}_k^T H(\mathbf{x}) \mathbf{u}_k + \phi(s) \label{subeq:clf-cbf-cost}\\
    \quad & \Delta V(\mathbf{x}_k, \mathbf{u}_k) + \alpha_k V(\mathbf{x}_k) \leq s \label{subeq:dclf-dcbf-clf-constraint}\\
    \quad & \Delta h(\mathbf{x}_k, \mathbf{u}_k) + \gamma_k h(\mathbf{x}_k) \geq 0 \label{subeq:dclf-dcbf-cbf-constraint}\\
    \quad & \mathbf{u}_k \in \mathcal{U},
\end{align}
\end{subequations}
\noindent\rule{\columnwidth}{0.4pt}
where $H(\mathbf{x})$ is any positive definite matrix and $s \geq 0$ is a slack variable together with an additional cost term $\phi(s) \geq 0$ that allows the Lyapunov function to grow when the CLF and CBF constraints are conflicting. The safe set $\mathcal{C}$ in \eqref{eq:cbf-safeset} is invariant along the trajectories of the discrete-time system with controller \eqref{eq:dclf-dcbf} if $h(\mathbf{x}_0)$ $\geq$ 0 and 0 $<$ $\gamma_k$ $\leq$ 1.

\subsubsection{MPC-DCBF~\cite{zeng2020safety}}
Inspired by the previous work of model predictive control and control barrier functions, DCLF-DCBF can be improved by taking future state prediction into account, yielding the form of MPC-DCBF firstly introduced in \cite{zeng2020safety}:

\noindent\rule{\columnwidth}{0.4pt}
\textbf{MPC-DCBF:}
\begin{subequations}
\label{eq:mpc-cbf}
\begin{align}
    J_{t}^{*}(\mathbf{x}_t){=}\min_{U} p(\mathbf{x}_{t+N|t}){+}\sum_{k=0}^{N-1}&q(\mathbf{x}_{t+k|t}, \mathbf{u}_{t+k|t}) \label{subeq:mpc-cbf-cost}\\
    \text{s.t.} \quad 
    \mathbf{x}_{t+k+1|t} = f(\mathbf{x}_{t+k|t}, \mathbf{u}_{t+k|t}), &\ k = 0,...,N{-}1\label{subeq:mpc-cbf-dynamics} \\
    \mathbf{u}_{t+k|t} \in \mathcal{U}, \ \mathbf{x}_{t+k|t} \in \mathcal{X}, &\ k = 0,...,N{-}1 \label{subeq:mpc-cbf-constraint} \\
    \mathbf{x}_{t|t} = \mathbf{x}_t, & \label{subeq:mpc-cbf-initial-condition} \\
    \Delta h (\mathbf{x}_{t+k|t}, \mathbf{u}_{t+k|t}) \geq -\gamma_k h(\mathbf{x}_{t+k|t}). &\ k = 0,...,N{-}1 \label{subeq:mpc-cbf-cbf}
\end{align}
\end{subequations}
\noindent\rule{\columnwidth}{0.4pt}
\noindent
Here, at each time $t$, the optimized cost-to-go function is denoted as $ J_{t}^{*}(\mathbf{x}_t)$ and $U = [\mathbf{u}_{t|t}^T, ..., \mathbf{u}_{t+N-1|t}^T]^T$.
In the cost function, $p(\mathbf{x}_{t+N|t})$ and $q(\mathbf{x}_{t+k|t}, \mathbf{u}_{t+k|t})$ represent terminal cost and stage cost at each time step.
The constraint \eqref{subeq:mpc-cbf-dynamics} describes the system dynamics, \eqref{subeq:mpc-cbf-constraint} shows the input constraints along the horizon and \eqref{subeq:mpc-cbf-initial-condition} represents the initial condition constraint.
The CBF constraints imposed in \eqref{subeq:mpc-cbf-cbf} are designed to guarantee the forward invariance of the safe set $\mathcal{C}$ associated with the discrete-time control barrier function.
Here we have
\begin{equation*}
    \Delta h (\mathbf{x}_{t+k|t}, \mathbf{u}_{t+k|t}) = h(\mathbf{x}_{t+k+1|t}) - h(\mathbf{x}_{t+k|t}).
\end{equation*}
The optimal solution to \eqref{eq:mpc-cbf} at time $t$ is a sequence of inputs as $U^{*} = [\mathbf{u}_{t|t}^{*T},...,\mathbf{u}_{t+N-1|t}^{*T}]^T$. Then, the first element of the optimizer vector is applied, i.e., 
\begin{equation}
\mathbf{u}_t = \mathbf{u}_{t|t}^{*}(\mathbf{x}_t). \label{eq:mpc-cbf-law}
\end{equation}
This constrained finite-time optimal control problem \eqref{eq:mpc-cbf} is repeated at time step $t+1$, based on the new state $\mathbf{x}_{t+1|t+1}$, yielding a receding horizon control strategy. 
In \cite{zeng2020safety}, MPC-DCBF has been shown to have better exploration performance with predictive horizon,  which can be beneficial in maneuvering through deadlock conditions.
However, MPC-DCBF has a smaller region of feasibility than the DCLF-DCBF controller as more constraints are used to confine the optimal control input.

\subsubsection{MPC-GCBF~\cite{ma2021feasibility}}
In the formulation \eqref{eq:mpc-cbf}, the CBF constraints are imposed on multi-steps along the horizon. This increases the computational complexity with a large horizon and the possibility of infeasibility also increases with more constraints.
One way to suppress the computational complexity and enhance the feasibility is to apply a one-step CBF constraint, which is done by modifying the CBF constraint in \eqref{subeq:mpc-cbf-cbf} as follows,
\begin{equation}
    \Delta h (\mathbf{x}_{t+1|t}, \mathbf{u}_{t|t}) \geq -\gamma h(\mathbf{x}_{t|t}). \label{eq:gcbf-one-order}
\end{equation}
This approach is firstly shown in the MPC-GCBF formulation in \cite{ma2021feasibility}, and enhances the feasibility and reduces the computational time at the same time due to fewer constraints.
This approach is also generalized with consideration over the high relative-degree constraint, where constraints in \eqref{subeq:mpc-cbf-cbf} are modified as constraints posed on two nonadjacent steps, 
\begin{equation}
    h (\mathbf{x}_{t+m|t}) \geq (1 - \gamma)^m h(\mathbf{x}_{t|t}), \label{eq:gcbf-high-order}
\end{equation}
where $m$ represents the relative degree of the high-order safety constraint.

\subsubsection{CLF-NMPC~\cite{grandia2020nonlinear}}
Beside the control barrier functions, the model predictive control is also unified with control Lyapunov functions, where stability constraints with CLFs are considered in the proposed CLF-NMPC formulation \cite{grandia2020nonlinear}.
The CLF constraint can be imposed during one-step or multi-step,
\begin{equation}
    \Delta V(\mathbf{x}_{t+k|t}) \leq -\alpha_k V(\mathbf{x}_{t+k|t}) + s_k,
\end{equation}
where $s_k$ is the slack variable.
The safety criteria is not considered in CLF-NMPC. For more details see ~\cite{grandia2020nonlinear}.

\section{CLFs and CBFs unified with NMPC}
\label{sec:formulation}
In this section, we are going to present two proposed formulations: NMPC-DCBF and NMPC-DCLF-DCBF unifying discrete-time CLFs and CBFs with NMPC. 

\subsection{Formulations}
\label{subsec:formulation}
Firstly, we incorporate discrete-time CBF constraints of decay rates under relaxing technique into a nonlinear model predictive control framework, denoted as NMPC-DCBF with the decision variables are $U = [\mathbf{u}_{t|t}^T, ..., \mathbf{u}_{t+N-1|t}^T]^T$ and $\Omega = [\omega_1, ..., \omega_{M_{\text{CBF}}-1}]^T$.

\noindent\rule{\columnwidth}{0.4pt}
\textbf{NMPC-DCBF:}
\begin{subequations}
\label{eq:cbf-nmpc}
\begin{align}
    J_{t}^{*}(\mathbf{x}_t){=}\min_{U, \Omega} \beta V(\mathbf{x}_{t+N|t}){+}\sum_{k=0}^{N-1}q(\mathbf{x}&_{t+k|t},\mathbf{u}_{t+k|t}){+}\psi(\omega_k) \label{subeq:cbf-nmpc-cost}\\
    \text{s.t.} \quad 
    \mathbf{x}_{t+k+1|t} = f(\mathbf{x}_{t+k|t}, \mathbf{u}_{t+k|t}), &\ k{=}0,...,N{-}1\label{subeq:cbf-nmpc-dynamics-constraint} \\
    \mathbf{u}_{t+k|t} \in \mathcal{U}, \ \mathbf{x}_{t+k|t} \in \mathcal{X},&\ k{=}0,...,N{-}1 \label{subeq:cbf-nmpc-input-constraint} \\
    \mathbf{x}_{t|t} = \mathbf{x}_t,& \label{subeq:cbf-nmpc-initial-condition} \\
    h (\mathbf{x}_{t+k+1|t}) \geq  \omega_k(1 -  \gamma_k) h(\mathbf{x}_{t+k|t}),& \ \omega_k \geq 0 \notag \\
     \text{for}\ k{=}0,...,&M_{\text{CBF}}{-}1 \label{subeq:cbf-nmpc-cbf-constraint}
\end{align}
\end{subequations}
\noindent\rule{\columnwidth}{0.4pt}
\noindent
The system dynamics \eqref{subeq:cbf-nmpc-dynamics-constraint}, the input constraint \eqref{subeq:cbf-nmpc-input-constraint} and the initial condition \eqref{subeq:cbf-nmpc-initial-condition} are imposed on the optimization.
The control Lyapunov function $V(\mathbf{x}_{t+N|t})$ is used as a terminal cost scaled up with the parameter $\beta$, together with the cumulative stage cost along the horizon $\sum_{k=0}^{N-1}q(\mathbf{x}_{t+k|t}, \mathbf{u}_{t+k|t})$. The terminal cost as CLF adopts the fashion of work from the field of MPC, as noted in \cite{rawlings2017model}, where stability usually can be achieved without the need to specify a terminal state constraint if $\beta$ is selected large enough.

Different from the formulation in \eqref{eq:mpc-cbf}, the decay rates of control barrier functions are relaxed from the fixed value $1 - \gamma_k$ into optimization variables $\omega_k(1 - \gamma_k)$ and an additional cost about the relaxing rate variables $\psi(\omega_k) \geq 0$ is included in the optimization. 
This function $\psi$ can be tuned for different performance.
The slack variable $\omega_k$ for relaxing is constrained by \eqref{subeq:cbf-nmpc-cbf-constraint} such that the following relation is guaranteed,
\begin{equation}
    h (\mathbf{x}_{t+k+1|t}) \geq  \omega_k(1 - \gamma_k) h(\mathbf{x}_{t+k|t}) \geq 0.
\end{equation}
This results in the safety guarantee for the first $M_{\text{CBF}}$ steps in the open-loop trajectory but not for the entire horizon $N$.
Here, one horizon length $N$ is designed for dynamics constraint, input constraint and stage cost, and another horizon length $M_{\text{CBF}} \leq N$ is applied for CBF constraints, which allows us to choose the appropriate value of $M_{\text{CBF}}$ to reduce the computational complexity.
We denote this formulation as NMPC-DCBF as the optimization is always nonlinear since the constraints \eqref{subeq:cbf-nmpc-cbf-constraint} are nonlinear even if system dynamics and the $h(.)$ function are linear. Note that the constraints inside a MPC-DCBF could become linear if the system dynamics and the $h(.)$ function are linear.

\begin{remark}
Here, we hypothesize that the closed-loop trajectory can still be guaranteed by iterations.
Formal guarantee of this property requires analysis of recursive feasibility and reachability, which will be proved in the subsequent work.
\end{remark}

\begin{remark}
NMPC-DCBF represents a generalized form of MPC-DCBF and MPC-GCBF with the relaxing technique of decay rates of safety constraints.
When the slack variable $\omega_k$ is fixed as 1, NMPC-DCBF becomes the same as MPC-DCBF when $M_{\text{CBF}} = N$ and the same as MPC-GCBF when $M_{\text{CBF}} = 1$.
\end{remark}

\begin{remark}
The fixed decay rates for safety constraints existing in MPC-DCBF and MPC-GCBF are relaxed and become as optimization variables in NMPC-DCBF, which increase the optimization feasibility.
\end{remark}

\begin{remark}
MPC-GCBF reduces the computational complexity and increases feasibility by reducing multi-step constraints into one-step.
However, one-step constraint might not confine the system sufficiently and the optimization problem may become infeasible after a while in the closed-trajectory, as shown in Fig. \ref{subfig:safety-mpcgcbf}.
Moreover, its set invariance for high-relative degree constraint relies on additional assumptions that $h(\mathbf{x}_{t+i|t}) > 0$ for $i = 0, 1,... m{-}1$, shown in \cite[Thm. 2]{ma2021feasibility}. This could be invalid as it depends on the initial condition of the system.
Additionally, identifying the high-relative degree of general complex dynamical systems could be difficult.
In our proposed approach, we continue to use the multi-step constraints, which guarantee stronger set invariance with less assumptions.
To apply NMPC-DCBF on a high-relative degree system, we just simply need $M_{\text{CBF}} \geq m$, where $m$ represents the relative-degree defined in \eqref{eq:gcbf-high-order}.
\end{remark}

Alternatively, the stability criteria with CLF could be posed as constraints instead of as a terminal cost, which leads to the formulation as follows,

\noindent\rule{\columnwidth}{0.4pt}
\textbf{NMPC-DCLF-DCBF:}
\begin{subequations}
\label{eq:clf-cbf-nmpc}
\begin{align}
    J_{t}^{*}(\mathbf{x}_t){=}\min_{U, \Omega, S} \sum_{k=0}^{N-1}q(\mathbf{x}_{t+k|t}, \mathbf{u}_{t+k|t})&{+}\psi(\omega_k){+}\phi(s_k)  \label{subeq:clf-cbf-nmpc-cost}\\
    \text{s.t.} \quad 
    \mathbf{x}_{t+k+1|t} = f(\mathbf{x}_{t+k|t}, \mathbf{u}_{t+k|t}), &\ k{=}0,...,N{-}1\label{subeq:clf-cbf-nmpc-dynamics-constraint} \\
    \mathbf{u}_{t+k|t} \in \mathcal{U},\ \mathbf{x}_{t+k|t} \in \mathcal{X}, &\ k{=}0,...,N{-}1 \label{subeq:clf-cbf-nmpc-input-constraint} \\
    \mathbf{x}_{t|t} = \mathbf{x}_t, & \label{subeq:clf-cbf-nmpc-initial-condition} \\
    h (\mathbf{x}_{t+k+1|t}) \geq  \omega_k(1 - \gamma_k) h(\mathbf{x}_{t+k|t}), & \ \omega_k \geq 0 \notag \\
    \text{for} \ k{=}0,...,&M_{\text{CBF}}{-}1 \label{subeq:clf-cbf-nmpc-cbf-constraint} \\
    V (\mathbf{x}_{t+k+1|t})\leq (1 - \alpha_k)V(\mathbf{x}_{t+k|t})&{+}s_k, \notag \\
    \text{for} \ k{=}0,...,&M_{\text{CLF}}{-}1 \label{subeq:clf-cbf-nmpc-clf-constraint}
\end{align}
\end{subequations}
\noindent\rule{\columnwidth}{0.4pt}
where $M_{\text{CLF}}$ and $M_{\text{CBF}}$ are the horizon length for CLF and CBF constraints respectively, which can be chosen to be less than the prediction horizon $N$. The slack variables $S = [s_1,..., s_{M_{\text{CLF}}}]^T$ are introduced to avoid infeasibility between CLF and CBF constraints and additional term $\phi(s_k) \geq 0$ is added into the cost function to minimize those slack variables.
This formulation is denoted as NMPC-DCLF-DCBF for later discussions.

\begin{remark}
NMPC-DCLF-DCBF represents a generalized form for DCLF-DCBF with horizon lengths for cost and constraints.
When the horizon lengths for CLF and CBF constraints equal to one, i.e., $M_{\text{CLF}} = M_{\text{CBF}} = 1$, the NMPC-DCLF-DCBF becomes exactly as DCLF-DCBF except the decay rate of CBF constraint is relaxed. This relaxation is necessary as it enhance feasibility.
Compared with NMPC-CLF in \cite{grandia2020nonlinear}, NMPC-DCLF-DCBF represents an extended form by adding safety constraints.
\end{remark}

\begin{remark}
On one hand, the proposed formulations \eqref{eq:cbf-nmpc}, \eqref{eq:clf-cbf-nmpc} have additional computational complexity generated due to the introduction of the optimization variables $\omega_k$. 
On the other hand, reducing the horizon length for constraints ($M_{\text{CBF}}$ and $M_{\text{CLF}}$) could reduce the computational complexity.
The joint influence on computational complexity arising from the additional optimization variables and reduced constraint horizons for the optimization problem depends on the system dynamics and non-linearity of the control barrier functions.
When the control barrier functions are nonlinear, the majority of non-linearity in the optimization comes from the CBF constraints, and therefore the reduction in complexity arising from the reduced constraint horizons would dominate the increase in complexity that arises from the introduction of additional optimization variables $\omega_k$.
\end{remark}
\section{Theoretical Analysis}
\label{sec:advantages}
In this section, we are going to illustrate theoretically the advantages about feasibility and safety of the proposed approaches.
These advantages compared with the state-of-the-art are summarized in Table \ref{tab:benchmark}.
\begin{table*}
\centering
\begin{tabular}{|c|cccc|cc|}
\hline
\multirow{2}{*}{\begin{tabular}[c]{@{}c@{}}Existing \& Proposed\\ Approaches\end{tabular}} & \multicolumn{4}{c|}{Optimization Structure} & \multicolumn{2}{c|}{Performance} \\ \cline{2-7} 
 & \multicolumn{1}{c|}{stability criteria} & \multicolumn{1}{c|}{safety criteria} & \multicolumn{1}{c|}{cost function} & constraint(s) & \multicolumn{1}{c|}{feasibility} & safety \\ \hline \hline
DCLF-DCBF~\cite{agrawal2017discrete} & constraint & constraint & one-step & one-step & medium & \bf{strong} \\ \hline
MPC-DCLF~\cite{grandia2020nonlinear} & constraints & none & multi-step & one/multi-step & \bf{high} & none \\ \hline
MPC-DCBF~\cite{zeng2020safety} & cost & constraints & multi-step & multi-step & low & \bf{strong} \\ \hline
MPC-GCBF~\cite{ma2021feasibility} & cost & constraint & multi-step & one-step & medium & weak \\ \hline \hline
NMPC-DCBF \eqref{eq:cbf-nmpc} & cost & constraint(s) & multi-step & one/multi-step & \bf{high} & \bf{strong} \\ \hline
NMPC-DCLF-DCBF \eqref{eq:clf-cbf-nmpc} & constraint(s) & constraint(s) & multi-step & one/multi-step & \bf{high} & \bf{strong} \\ \hline
\end{tabular}
\caption{A comparison among existing and proposed optimal control methods with respect to a variety of attributes.}
\label{tab:benchmark}
\end{table*}

\subsection{Theoretical Analysis of Feasibility}
\label{subsec:feasibility}
In this section, we are going to illustrate the enhancement of feasibility with reachability analysis by comparing MPC-DCBF and NMPC-DCBF.

For the MPC-DCBF formulation, the reachable set and safe region confined by the CBF constraint at each time are defined respectively as follows:
\begin{equation}
\label{eq:reachable-set-mpc-cbf}
\begin{split}
    \mathcal{R}^{\text{MPC-DCBF}}_k = \{\mathbf{x}_{t+k|t} \in \mathbb{R}^{n}: \forall i = 0,...,k-1, \\
    \mathbf{x}_{t+i+1|t} = f(\mathbf{x}_{t+i|t}, \mathbf{u}_{t+i|t}), \\ 
    \mathbf{u}_{t+i|t} \in \mathcal{U}, \mathbf{x}_{t+k|t} \in \mathcal{X}, \mathbf{x}_{t|t} = \mathbf{x}_t\},
\end{split}
\end{equation}

\begin{equation}
\label{eq:safe-region-mpc-cbf-original}
\begin{split}
    \mathcal{S}^{\text{MPC-DCBF}}_{k} = \{ & \mathbf{x}_{t+k|t} \in \mathbb{R}^{n} : h(\mathbf{x}_{t+k|t})\\
    &  {-}h(\mathbf{x}_{t+k-1|t}) {\geq} {-}\gamma_k h (\mathbf{x}_{t+k-1|t}) \}.
\end{split}
\end{equation}
The optimization of MPC-DCBF is feasible when the intersections between the reachable set $\mathcal{R}^{\text{MPC-DCBF}}_{k}$ and safe set $\mathcal{S}^{\text{MPC-DCBF}}_{k}$ at time $t+k$ are non-empty for all $k$.
We denote the safe region at each time step as $\mathcal{S}^{\text{MPC-DCBF}}_{k}$, but notice that it also depends on the value of optimal value $\mathbf{x}_{t+k-1|t}$, which depends on the states and the inputs of previous nodes before the index $k-1$.
$\mathcal{S}^{\text{MPC-DCBF}}_{k}$ could be rewritten as a function of reachable set at the one-step before,
\begin{equation}
\label{eq:safe-region-mpc-cbf}
\begin{split}
    \mathcal{S}^{\text{MPC-DCBF}}_{k} = &\{ \mathbf{x}_{t+k|t} \in \mathbb{R}^{n} :\\
    & h(\mathbf{x}_{t+k|t}) \geq (1 - \gamma_k) \inf_{\mathbf{x} \in \mathcal{R}^{\text{MPC-DCBF}}_{k-1}} h(\mathbf{x}) \}, \\
\end{split}
\end{equation}
as \eqref{eq:safe-region-mpc-cbf-original} leads to the following equation being valid
\begin{equation*}
    h(\mathbf{x}_{t+k|t}) \geq (1-\gamma_k) h(\mathbf{x}_{t+k-1|t}),
\end{equation*}
with at least one value of $\mathbf{x}_{t+k-1|t}$.

For the NMPC-DCBF formulation, the reachable set is the same as MPC-DCBF as they share the same initial condition, system dynamics and input constraints, i.e.,
\begin{equation*}
    \mathcal{R}^{\text{NMPC-DCBF}}_k = \mathcal{R}^{\text{MPC-DCBF}}_k.
\end{equation*}
The corresponding safe region at each step is as follows,
\begin{equation}
\label{eq:safe-region-cbf-nmpc-original}
\begin{split}
    &\mathcal{S}^{\text{NMPC-DCBF}}_{k} = \{ \mathbf{x}_{t+k|t} \in \mathbb{R}^{n} : \omega_k \geq 0, \\
    & h(\mathbf{x}_{t+k|t}) \geq \omega_k(1 - \gamma_k) \inf_{\mathbf{x} \in \mathcal{R}^{\text{NMPC-DCBF}}_{k-1}} h(\mathbf{x}) \}. \\
\end{split}
\end{equation}
As $\omega_k$ is an optimization variable with constraint $\omega_k \geq 0$, this leads us to rewrite $\mathcal{S}^{\text{NMPC-DCBF}}_{k}$ as follows,
\begin{equation}
\label{eq:safe-region-cbf-nmpc}
\begin{split}
    &\mathcal{S}^{\text{NMPC-DCBF}}_{k} = \{ \mathbf{x}_{t+k|t} \in \mathbb{R}^{n} : h(\mathbf{x}_{t+k|t}) \geq 0 \} = \mathcal{C}. \\
\end{split}
\end{equation}
We can see that any value of $\gamma_k$ for NMPC-DCBF won't affect the feasibility, as $\mathcal{S}^{\text{NMPC-DCBF}}_{k}$ equals the safe set $\mathcal{C}$ defined in \eqref{eq:cbf-safeset}, which is independent of $\gamma_k$.
Hence, we have $\mathcal{S}^{\text{MPC-DCBF}}_{k}$ in \eqref{eq:safe-region-mpc-cbf} always as a subset of $\mathcal{S}^{\text{NMPC-DCBF}}_{k}$ in \eqref{eq:safe-region-cbf-nmpc}, as
\begin{equation}
    \mathcal{S}^{\text{MPC-DCBF}}_{k} \subset \mathcal{S}^{\text{NMPC-DCBF}}_{k} = \mathcal{C}
\end{equation}
which results to show that the feasible regions of decision variable $\mathbf{x}_{t+k|t}$ is always larger when applying NMPC-DCBF approach compared to MPC-DCBF
\begin{equation}
    \mathcal{R}^{\text{MPC-DCBF}}_k \cap \mathcal{S}^{\text{MPC-DCBF}}_{k} \subset \mathcal{R}^{\text{NMPC-DCBF}}_k \cap \mathcal{S}^{\text{NMPC-DCBF}}_{k},
\end{equation}
where feasibility region at step $k$ is the intersection between reachable set $\mathcal{R}_k$ and safe set $\mathcal{S}_k$ for both approaches.
To sum up, as the relaxing technique for decay rates is applied in NMPC-DCBF and NMPC-DCLF-DCBF, we can state they outperform MPC-DCBF / MPC-GCBF / DCLF-DCBF from the perspective of feasibility.

\begin{remark}
In \eqref{eq:safe-region-cbf-nmpc}, it shows that $\mathcal{S}^{\text{NMPC-DCBF}}_{k} = \mathcal{C}$, which is the same as the region enforced by the distance constraint \eqref{eq:distance-constraint}.
This reveals the NMPC-DCBF holds the same feasible region as MPC-DC~\cite{zeng2020safety}.
Therefore, the feasible regions for NMPC-DCBF and NMPC-DCLF-DCBF are constant with respect to different hyperparameters $\gamma_k$, which will be shown in Fig. \ref{fig:feasibility-mpccbf-cbfnmpc}, \ref{fig:feasibility-mpcgcbf-cbfnmpc} and \ref{fig:feasibility-dclfdcbf-clfcbfnmpc}.
\end{remark}

\subsection{Theoretical Analysis of Safety}
\label{subsec:safety}
The system safety can be influenced by many factors, including the safety function, the cost function, and other hyperparameters, etc.
In this section, we focus on the influence from the hyperparameter $\gamma_k$ and the additional cost function $\psi(\omega_k)$ for the slack variable $\omega_k$ of decay rate.

By reducing $\gamma_k$ for MPC-DCBF / MPC-GCBF / DCLF-DCBF, the system safety will increase as the smaller $\gamma_k$ represents a slower decay rate of lower bound of control barrier function, see \eqref{eq:cbf-definition}.
However, from \eqref{eq:safe-region-mpc-cbf}, we can see that reducing $\gamma_k$ for MPC-DCBF makes $\mathcal{S}^{\text{MPC-DCBF}}_k$ smaller, which leads the optimization more likely to be infeasible along the trajectory as the intersection between the reachable set and the region constrained by safety constraint decreases.
This leads to a tradeoff between feasibility and safety.
This tradeoff also happens among DCLF-DCBF and MPC-GCBF with similar reasons and forces us to choose either feasibility or safety for performance.
However, in the case of our proposed NMPC-DCBF, the region confined by safety constraint won't be affected by changing the value of $\gamma_k$, as shown in \eqref{eq:safe-region-cbf-nmpc}.
Hence, the intersection between the reachable set and the region constrained by safety constraint is independent of $\gamma$.
This allows us to enhance the safety by reducing $\gamma_k$ while not harming feasibility, which resolves the tradeoff between feasibility and safety.

The design of the additional cost function $\psi(\omega_k)$ for the decay-rate slack variable $\omega_k$ could also affect the safety performance. For example, the function $\psi(\omega_k)$ can be in the form as follows,
\begin{equation}
    \psi(\omega_k) = P_{\omega}(\omega_k - 1)^2
\end{equation}
which keeps $\omega_k$ close to 1 and thus minimizes the deviation of the CBF constraint from the nominal decay rate of $1-\gamma_k$.
When the hyperparameter $P_{\omega}$ becomes larger, the optimized value of $\omega$ tends to be closer to 1, which implies the deviation from the nominal decay rate $1-\gamma_k$ is smaller.
Numerically, $\gamma_k$ tends to be optimized as value smaller than 1 to increase the safe region \eqref{eq:safe-region-cbf-nmpc} confined by CBF constraint at each time step.
When $\omega_k = 0$, the relaxed CBF constraint becomes equivalent to a simple distance constraint and MPC with distance constraints (MPC-DC~\cite{zeng2020safety}) needs longer horizon to generate an expected obstacle avoidance performance in a closed-loop trajectory.
Therefore, it's not recommended to set a relatively too small value for $P_{\omega}$, which would over-relax the CBF constraint, i.e., the optimized value of $\omega_k$ could be too small.

To sum up, by reducing $\gamma_k$ and utilizing an appropriate form of the additional cost function for decay-rate slack variable $\omega_k$, the proposed approach would outperform the existing formulations in term of safety while not harming feasibility performance.

\section{Numerical Examples \& Results}
\label{sec:examples}
\begin{figure*}[!htp]
    \centering
    \begin{subfigure}[t]{0.23\linewidth}
        \centering
        \includegraphics[width = 0.99\linewidth]{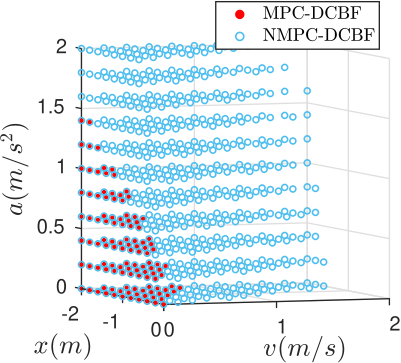}\\
        \caption{$\gamma_k = 0.05$}
        \label{subfig:feasibility-mpccbf1}
    \end{subfigure}
    \begin{subfigure}[t]{0.23\linewidth}
        \centering
        \includegraphics[width = 0.99\linewidth]{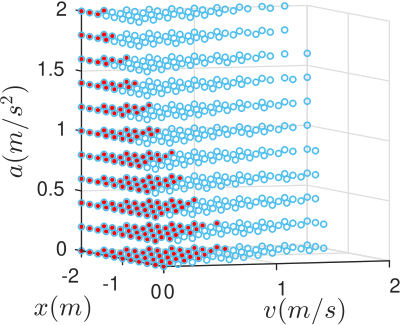}\\
        \caption{$\gamma_k = 0.10$}
        \label{subfig:feasibility-mpccbf2}
    \end{subfigure}
    \begin{subfigure}[t]{0.23\linewidth}
        \centering
        \includegraphics[width = 0.99\linewidth]{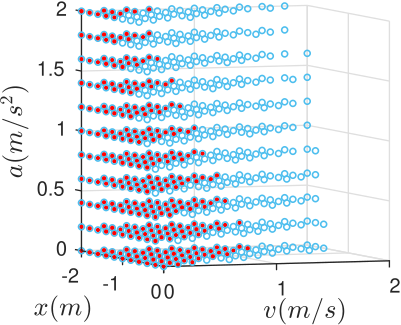}\\
        \caption{$\gamma_k = 0.15$}
        \label{subfig:feasibility-mpccbf3}
    \end{subfigure}
    \begin{subfigure}[t]{0.23\linewidth}
        \centering
        \includegraphics[width = 0.99\linewidth]{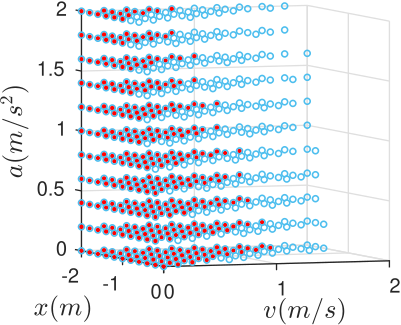}\\
        \caption{$\gamma_k = 0.20$}
        \label{subfig:feasibility-mpccbf4}
    \end{subfigure}
    \caption{Feasibility comparison with $h(\mathbf{x}) = -x$ between MPC-DCBF ($N = 8$) and NMPC-DCBF ($N = 8, M_{\text{CBF}} = 8$) with different values of $\gamma_k$.
    Feasible states are marked by red points (MPC-DCBF) and blue circles (NMPC-DCBF).
    It's shown that the feasibility region of MPC-DCBF is always a subset of feasibility region of NMPC-DCBF, and the feasibility region of NMPC-DCBF is independent of $\gamma_k$.}
    \label{fig:feasibility-mpccbf-cbfnmpc}
\end{figure*}
\begin{figure*}[!htp]
    \centering
    \begin{subfigure}[t]{0.23\linewidth}
        \centering
        \includegraphics[width = 0.99\linewidth]{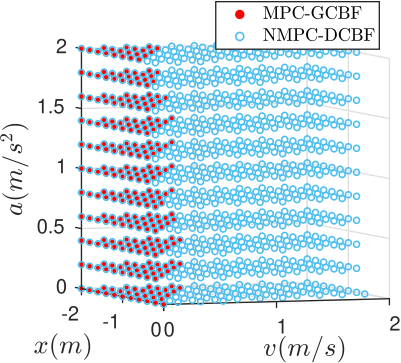}\\
        \caption{$\gamma_k = 0.05$}
        \label{subfig:feasibility-mpcgcbf1}
    \end{subfigure}
    \begin{subfigure}[t]{0.23\linewidth}
        \centering
        \includegraphics[width = 0.99\linewidth]{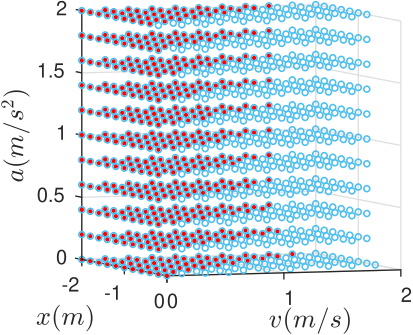}\\
        \caption{$\gamma_k = 0.10$}
        \label{subfig:feasibility-mpcgcbf2}
    \end{subfigure}
    \begin{subfigure}[t]{0.23\linewidth}
        \centering
        \includegraphics[width = 0.99\linewidth]{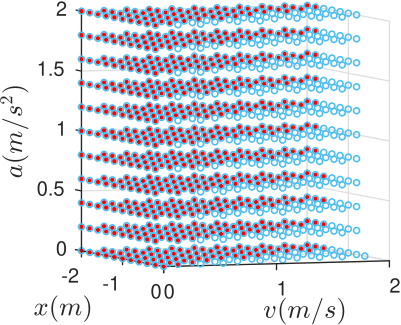}\\
        \caption{$\gamma_k = 0.15$}
        \label{subfig:feasibility-mpcgcbf3}
    \end{subfigure}
    \begin{subfigure}[t]{0.23\linewidth}
        \centering
        \includegraphics[width = 0.99\linewidth]{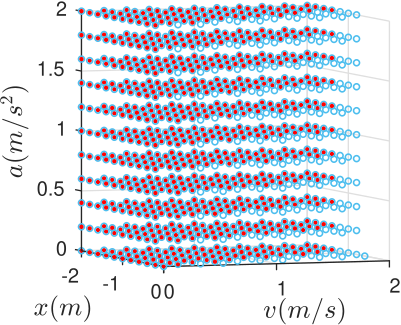}\\
        \caption{$\gamma_k = 0.20$}
        \label{subfig:feasibility-mpcgcbf4}
    \end{subfigure}
    \caption{Feasibility comparison with $h(\mathbf{x}) = -x$ between MPC-GCBF ($N = 8$) and NMPC-DCBF ($N = 8, M_{\text{CBF}} = 3$) with different values of $\gamma_k$.
    It's shown that the feasibility region of MPC-GCBF is always a subset of feasibility region of NMPC-DCBF, and the feasibility region of NMPC-DCBF is independent of $\gamma_k$.
    }
    \label{fig:feasibility-mpcgcbf-cbfnmpc}
\end{figure*}
\begin{figure*}[!htp]
    \centering
    \begin{subfigure}[t]{0.23\linewidth}
        \centering
        \includegraphics[width = 0.99\linewidth]{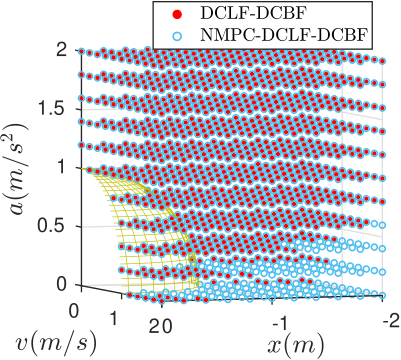}\\
        \caption{$\gamma_k = 0.05$}
        \label{subfig:feasibility-dclfdcbf1}
    \end{subfigure}
    \begin{subfigure}[t]{0.23\linewidth}
        \centering
        \includegraphics[width = 0.99\linewidth]{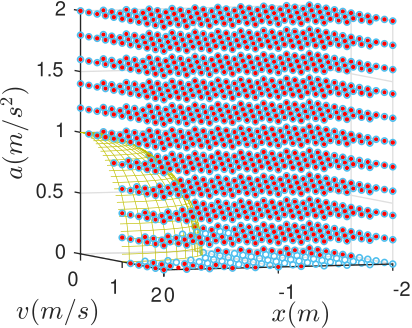}\\
        \caption{$\gamma_k = 0.10$}
        \label{subfig:feasibility-dclfdcbf2}
    \end{subfigure}
    \begin{subfigure}[t]{0.23\linewidth}
        \centering
        \includegraphics[width = 0.99\linewidth]{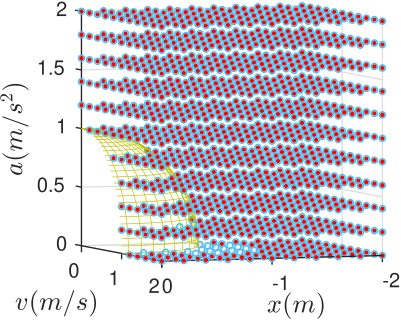}\\
        \caption{$\gamma_k = 0.15$}
        \label{subfig:feasibility-dclfdcbf3}
    \end{subfigure}
    \begin{subfigure}[t]{0.23\linewidth}
        \centering
        \includegraphics[width = 0.99\linewidth]{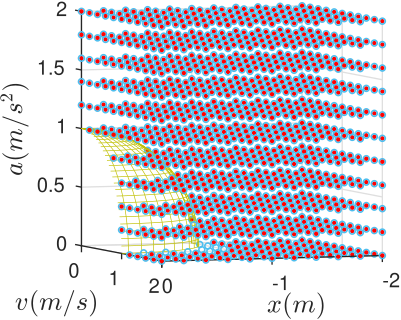}\\
        \caption{$\gamma_k = 0.20$}
        \label{subfig:feasibility-dclfdcbf4}
    \end{subfigure}
    \caption{Feasibility comparison with $h(\mathbf{x}) = x^2 + v^2 + a^2 - 1$ between DCLF-DCBF and NMPC-DCLF-DCBF ($N = 8, M_{\text{CLF}} = 8, M_{\text{CBF}} = 8$) with different values of $\gamma_k$.
    The zero-level set of CBF constraint is marked in yellow.
    It's shown that the feasibility region of DCLF-DCBF is always a subset of the one of NMPC-DCLF-DCBF, and the one of NMPC-DCLF-DCBF is independent of $\gamma_k$.}
    \label{fig:feasibility-dclfdcbf-clfcbfnmpc}
\end{figure*}
In this section, we are going to show numerical results to illustrate the advantages of our proposed formulations with respect to the existing approaches.
Consider the discrete-time linear triple-integrator system,
\begin{equation}
    \mathbf{x}_{k+1} = A \mathbf{x}_{k} + B \mathbf{u}_{k}
\end{equation}
where $\mathbf{x} = [x, v, a]^T$ and $\mathbf{u} = [j]^T$ represent position ($x$), velocity ($v$), acceleration ($a$) and jerk ($j$), respectively.
The admissible input set is $\mathcal{U} = \{j \in \mathbb {R}: j_{\min} \leq j \leq j_{\max} \}$.

For numerical simulations in this section, we set the sampling time as $\Delta t = 0.1 s$ together with input lower and upper bounds as $j_{\min,\max} = -1 m/s^3, 1 m/s^3$.
All simulations run in MATLAB and the optimal control is formulated with Yalmip~\cite{lofberg2004yalmip} as modelling language and solved with IPOPT~\cite{wachter2006implementation}.

\subsection{Numerical Results for Feasibility}
\label{subsec:feasibility-results}
Our proposed formulations along with existing ones is compared by solving the optimization problems at all sampling states in a closed space.
Precisely, we iterate over sampling states in the closed space $\mathcal{X}$ as 
\begin{equation*}
\begin{split}
    \mathcal{X} = \{ (x, v, a) \in \mathbb{R}^3 ~:~ & x_{\min} \leq x \leq x_{\max}, v_{\min} \leq v \leq v_{\max}, \\ 
    & a_{\min} \leq a \leq a_{\max} \}
\end{split}
\end{equation*}
and run these optimal controllers to see whether the optimization problems are feasible at a given state $\mathbf{x}_{t}$.
For simulation, we set $x_{\min, \max} = -2m, 0m$, $v_{\min, \max} = 0m/s, 2m/s$ and $a_{\min, \max} = 0m/s^2, 2m/s^2$.
All the feasibility performance comparison is evaluated between approaches with the same horizon $N$ and same form of stage cost and terminal cost.

\begin{figure*}
    \centering
    \begin{subfigure}[t]{0.32\linewidth}
        \centering
        \includegraphics[width = 0.99\linewidth]{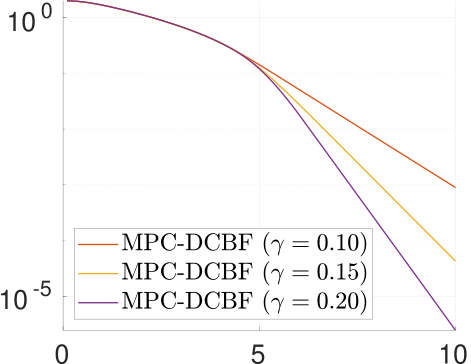}\\
        \caption{MPC-DCBF ($N = 8$)}
        \label{subfig:safety-mpccbf}
    \end{subfigure}
    \begin{subfigure}[t]{0.32\linewidth}
        \centering
        \includegraphics[width = 0.99\linewidth]{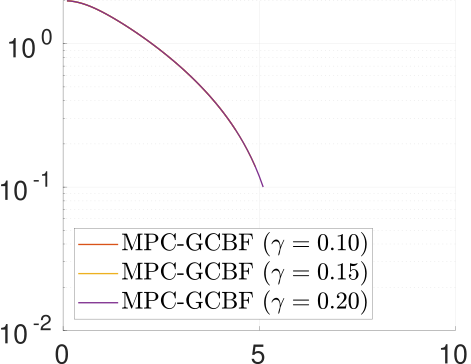}\\
        \caption{MPC-GCBF ($N = 8$)}
        \label{subfig:safety-mpcgcbf}
    \end{subfigure}
    \begin{subfigure}[t]{0.32\linewidth}
        \centering
        \includegraphics[width = 0.99\linewidth]{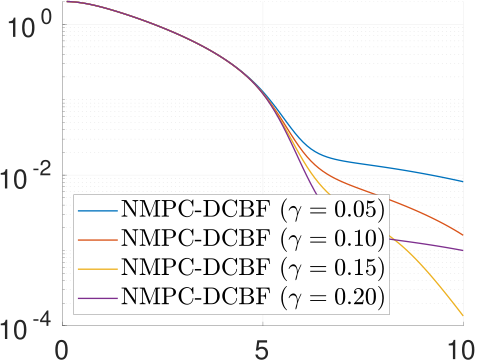}\\
        \caption{NMPC-DCBF ($N = 8, M_{\text{CBF}} = 8$)}
        \label{subfig:safety-cbfnmpc}
    \end{subfigure}
    \caption{Evolution of control barrier function $h(\mathbf{x}) = -x$ in the closed-loop trajectory by using controllers MPC-DCBF, MPC-GCBF and NMPC-DCBF with different values of $\gamma_k$. Notice that three evolution lines for MPC-GCBF in Fig. \ref{subfig:safety-mpcgcbf} overlap with each other and they all become infeasible around time $t = 5.0s$}. NMPC-DCBF is the only one to have feasibility when $\gamma = 0.05$.
    \label{fig:safety}
\end{figure*}

To compare the feasibility performance among MPC-DCBF, MPC-GCBF and NMPC-DCBF, we choose a high-order control barrier function
\begin{equation}
    h(\mathbf{x}) = -x, \label{eq:example-high-order-cbf}
\end{equation}
which enforces the system to stay on one side of the yz plane $(x \leq 0)$.
In Figs. \ref{fig:feasibility-mpccbf-cbfnmpc}, \ref{fig:feasibility-mpcgcbf-cbfnmpc}, results from NMPC-DCBF are compared with MPC-DCBF and MPC-GCBF for feasibility analysis.
The comparisons are validated with different values of $\gamma_k = 0.05, 0.10, 0.15, 0.20$.
For a reasonable comparison, the horizon length of CBF constraints is assumed as $M_{\text{CBF}} = 3$ for NMPC-DCBF to compare with MPC-GCBF, as the relative-degree of the CBF in \eqref{eq:example-high-order-cbf} is $3$ for a triple integrator system. 
The formulations of MPC-DCBF and MPC-GCBF are shown to enhance the feasibility with larger value of $\gamma$.
MPC-GCBF does enhance the feasibility compared with MPC-DCBF as more states are feasible for MPC-GCBF with any value of $\gamma_k$.
The feasibility of the proposed NMPC-DCBF is shown to consistently outperform MPC-DCBF and MPC-GCBF for any value of $\gamma_k$, where the feasible state region for MPC-DCBF or MPC-GCBF lies always inside the one for NMPC-DCBF.
Additionally, the feasible state region of NMPC-DCBF is independent of the value of $\gamma_k$, shown in Figs. \ref{fig:feasibility-mpccbf-cbfnmpc}, \ref{fig:feasibility-mpcgcbf-cbfnmpc}, which verifies $\mathcal{S}^{\text{NMPC-DCBF}}_{k}$ is independent of $\gamma_k$ as shown in \eqref{eq:safe-region-cbf-nmpc}.

To compare the feasibility performance between DCLF-DCBF and NMPC-DCLF-DCBF, we
choose a relative-degree one control barrier function
\begin{equation}
    h(\mathbf{x}) = x^2 + v^2 + a^2 -1, \label{eq:example-one-order-cbf}
\end{equation}
as the DCLF-DCBF method can only optimize one-step control input.
The comparison result is shown in Fig. \ref{fig:feasibility-dclfdcbf-clfcbfnmpc}, where NMPC-DCLF-DCBF outperforms DCLF-DCBF in terms of feasibility for any values of $\gamma_k$.
Similar to what we have seen previously, DCLF-DCBF enhances the feasibility with larger $\gamma_k$, while the feasible state region for NMPC-DCLF-DCBF is independent of $\gamma_k$. 
Notice that the unsafe states, which are inside the sphere region defined by \eqref{eq:example-one-order-cbf}, are excluded from the state sampling test for feasibility. The zero-level surface of the control barrier function is colored in yellow in Fig. \ref{fig:feasibility-dclfdcbf-clfcbfnmpc}.

We also remark that the number of safety constraints for NMPC-DCBF and NMPC-DCLF-DCBF are larger than the ones for MPC-GCBF and DCLF-DCBF, but the feasibility performance are enhanced in our proposed approach, which demonstrates the importance of decay-rate relaxing technique that are introduced in the two proposed formulations.
To sum up, the proposed formulations outperform the state-of-the-art in terms of feasibility.

\subsection{Safety}
\label{subsec:safety-results}
The safety performance between controllers are compared numerically in this section.
Given the same initial condition $\mathbf{x}_0 = [-2.0, 0.0, 1.0]^T$, we test each controller performance by using hyperparameters $\gamma_k = 0.05, 0.10, 0.15, 0.20$.
The results for comparison among these approaches are shown in Fig. \ref{fig:safety}.
Among MPC-DCBF and NMPC-DCBF, it can seen that by reducing the value of $\gamma_k$, the value of CBF decreases slower, which implies a safer closed-loop trajectory.
We notice that NMPC-DCBF is the only approach that maintains feasibility along the trajectory with $\gamma = 0.05$, while the other two approaches are infeasible right at the initial condition.
Additionally, as illustrated in Fig. \ref{subfig:safety-mpcgcbf}, 
MPC-GCBF becomes infeasible after around 5 seconds in a closed-loop trajectory starting from the initial condition.
This arises from the fact that the one-step constraint doesn't sufficiently confine the system for safety and leads the system into an infeasible state after a while.
We also notice that the control barrier function for NMPC-DCBF with $\gamma_k = 0.20$ is larger than $\gamma_k = 0.15$ after $t = 8s$. This happens due to numerical errors as
after $t = 8s$, the CBF $h(\mathbf{x})$ is very close to zero and its derivative becomes almost zero and the solver tends to optimize the additional cost for relaxing decay-rate variable instead of the stage and terminal cost.
Together with feasibility analysis in Sec. \ref{subsec:feasibility-results}, we have shown that by reducing $\gamma_k$, NMPC-DCBF could enhance the safety of the closed-loop trajectory while not adversely affecting feasibility, which resolves the tradeoff between feasibility and safety.

\section{Conclusion \& Future Work}
\label{sec:conclusion}
In this paper, we have proposed formulations to unify control Lyapunov function and control barrier functions under the framework of nonlinear model predictive control. Compared with previous work, the decay-rate of the CBF constraints are relaxed and different horizon lengths are considered for the cost function and the constraints.
The proposed formulations are shown both theoretically and numerically to outperform the state-of-the-art from the perspective of feasibility and safety.

Our future work will focus on how to implement the proposed formulations either as a mid-level planner or a real-time controller on mobile robots, where modelling uncertainty, system disturbance and noise are required to be considered for real-time deployment.
From the theoretical perspective, a formal discussion about recursive feasibility and stability will be carried out to summarize the technique of nonlinear MPC with control barrier function.
\balance
\bibliographystyle{IEEEtran}
\bibliography{references}{}

\end{document}